\documentclass[12pt,preprint]{aastex}

\begin{document}

\title{A second neutron star in M4?{\LARGE$^\ast$}
      }


\author{J. Kaluzny, A. Rozanska, M. Rozyczka, W. Krzeminski}
\affil{Nicolaus Copernicus Astronomical Center, Bartycka 18, 00-716 Warsaw, Poland}

\and
\author{Ian B. Thompson}
\affil{Observatories of the Carnegie Institution of Washington, 813 Santa Barbara Street, Pasadena, CA 91101}

\begin{abstract}
We show that the optical counterpart of the X-ray source CX 1 in M4 is a 
$\sim$20th magnitude star, located in the color-magnitude diagram on (or very close to) 
the main sequence of the cluster, and exhibiting sinusoidal variations of the flux. 
We find the X-ray flux to be also periodically variable, with X-ray and optical minima 
coinciding. Stability of the optical light curve, lack of UV-excess, and unrealistic
mean density resulting from period-density relation for semidetached systems, speak 
against the original identification of CX 1 as a cataclysmic variable. We argue that 
the X-ray active component of this system is a neutron star (probably a millisecond 
pulsar).  

\end{abstract}

\keywords{binaries: close -- globular clusters: general -- globular clusters: 
individual (M4) -- cataclysmic variables -- X-rays: stars}

\section{Introduction}
Globular clusters (GCs) contain a large population of exotic binary stars with degenerate
components, and many of these are X-ray sources with various degrees of activity. 
Numerous such sources have been detected by the
{\it Chandra} space observatory in over 80 distinct clusters, providing a valuable tracer 
population for large-scale numerical simulations of cluster evolution \citep{poo10}. 
\let\thefootnote\relax\footnotetext      
{$^\ast$Based on observations made with the NASA/ESA Hubble Space Telescope,
and obtained from the Hubble Legacy Archive, which is a collaboration
between the Space Telescope Science Institute (STScI/NASA), the Space
Telescope European Coordinating Facility (ST-ECF/ESA) and the
Canadian Astronomy Data Centre (CADC/NRC/CSA).}

One of the clusters surveyed by {\it Chandra} is M4 (NGC 6121). 
Eight of the X-ray sources found in its 
core region were classified as chromospherically active binaries, one as a cataclysmic variable 
(CV), and one was identified with the millisecond pulsar PSR1620-26 
\citep[][hereafter BPH]{bas04}. The CV-candidate, 
referred to as CX~1 in BPH, is the brightest among them 
($L_{0.5-2.5\mathrm{keV}}=8.3\times10^{31}$ erg~s$^{-1}$). BPH identified it with 
a bright $V_\mathrm{555}=17.37$ star, located on the main sequence of M4 just below the turnoff. 
They argued that its X-ray luminosity is too 
high for a magnetically active binary composed of two main-sequence stars (i.e., a BY Dra system), 
while its optical luminosity is too low for an active binary with a subgiant (i.e., an RS CVn 
system). Their identification of CX 1 as a CV was additionally supported by its rather hard X-ray 
spectrum ($L_{0.5-1.5\mathrm{keV}}/L_{1.5-6.0 \mathrm{keV}}=1.22$), and a high ratio of X-ray and 
optical fluxes. 

BPH noted that the optical centroid of the bright star is offset by 2$\sigma$ from the 
{\it Chandra} position of CX 1, so the identification might be questionable. If 
that were the case, 
the actual counterpart of CX 1 could be much fainter, possibly lost in the glare of the brighter 
star. Indeed, the examination of the M4-photometry published by \citet{and08} showed that 
the object identified by BPH was a blend of 
two stars. In the present Letter we demonstrate that the X-ray activity is 
associated with the much fainter and redder component of the blend. We argue that CX 1 is a
millisecond pulsar or qLMXB rather than a cataclysmic variable.

\section{Observations}
\label{sect:obser}

\subsection{Ground-based}
\label{sect:lascam}
M4 was observed in nine seasons between 1995 and 2009 as one of the objects included in 
the CASE project \citep{jka05}. Altogether, 1683 $V$-band images were collected 
with the 2.5-m du Pont telescope at Las Campanas Observatory, using the same CCD camera 
and the same filter. 
Exposure times ranged  from  20s to 120s, with a median value of 45s. The median 
seeing was $0.98^{\prime\prime}$. The photometric analysis was based on a combination 
of image subtraction and profile photometry techniques. We used scripts 
employing the Daophot/Allstar package \citep{ste87} and a modified version of the 
ISIS code \citep{ala98}; see \citet{jka10} for details. Using ISIS, one obtains a light
curve in the form of differential counts. This allows for the photometry of variable 
objects even in those cases when they are tightly blended.  
Conversion of the light curve from differential counts to stellar magnitudes
requires determination of the pedestal flux for a target star on the reference
image (see Sect. \ref{sect:hst}). 
Further details concerning 
the observations and the calibration of the photometry will be given in a forthcoming 
paper (Kaluzny et al. 2012; in preparation).

\begin{figure}[ht!]
\centering
\includegraphics[width=0.46\textwidth,bb= 29 167 404 690,clip]{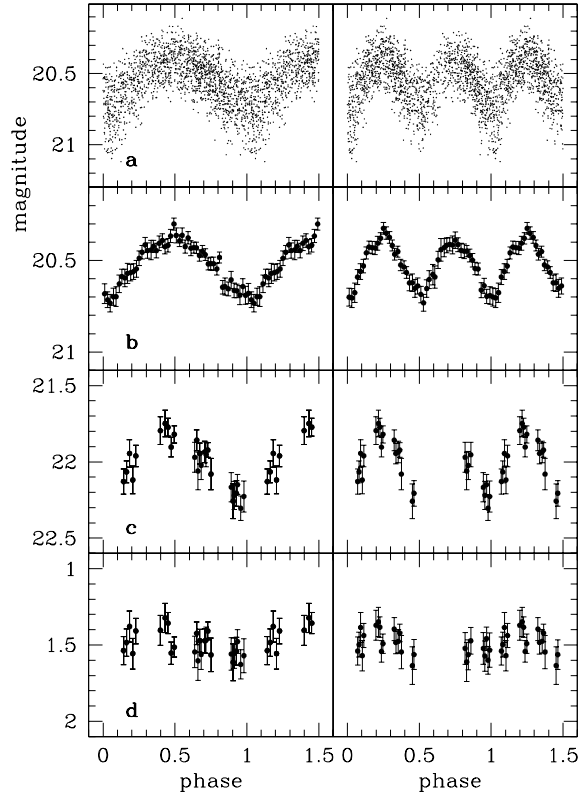}
\caption {Left column: optical light curves of the star \#5274 phased with the ephemeris 
          (\ref{eq:ephem}). a: ground-based, $V$-band (mean error of a single 
           observation is 0.20 mag); b: the same data binned (50 bins per period);
           c: archival HST data, F435W-band, ACS/VEGA system; 
           d: color index (F435W - $V$). Right column: the same data phased with 
           $P_\mathrm{2}$ used in eq. (\ref{eq:ephem}) instead of $P_\mathrm{1}$.
         }
\label{fig:opt_lc}
\end{figure}

The optical counterpart of CX 1 proposed by BPH turned out to be a 
low-amplitude periodic variable with a stable light curve, suggestive of an ellipsoidal
and/or irradiation effect. The analysis of variance, implemented in the $Tatry$ code 
\citep{asc96}, yielded a period $P_\mathrm{1}=0.2628216(1)$ d. The linear 
ephemeris 
\begin{equation}
 HJD_\mathrm{min} = 2,454,656.7394(16) + E\times P_\mathrm{1} 
 \label{eq:ephem}
\end{equation}
provides a good fit to all data - we found no evidence for any change of period, 
amplitude or average magnitude from one observational season to another.
There is, however, an ambiguity concerning the period: a value 
$P_\mathrm{2} = 0.5256432(1)$~d, i.e. two times larger than $P_\mathrm{1}$, 
is also perfectly acceptable (see Fig. \ref{fig:opt_lc}). We address this issue in 
Sects.~\ref{sect:hst}~and~\ref{sect:discus}.

\subsection{HST}
\label{sect:hst}
The optical counterpart of CX 1 is visible not only in {\it HST/WFPC2} images analysed by 
BPH, but also in several images obtained with {\it HST/ACS}. A closer inspection 
of those data reveals that it is actually a pair of stars, separated by $0.15^{\prime\prime}$. 
These stars were resolved in the F606W/F814W photometry performed by \citet{and08}, and placed 
in their data base with F606W, F814W, $V$ and $I$ magnitudes. According to \citet{and08}, the 
brighter component of the pair (star \#5273) has $V=17.170$, while the fainter one (star \#5274)
reaches only $V=20.650$. We note that the difference between $V=17.170$ of \citet{and08} and 
$V_\mathrm{555}=17.37$ of BPH originates from different photometric systems used by those authors. 

\begin{figure}[t!]
\centering
\includegraphics[width=0.46\textwidth, bb= 57 177 477 515,clip]{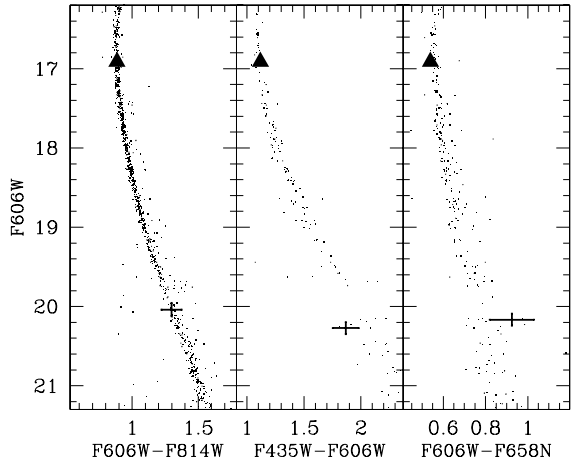}
\caption {Location of stars \#5273 (triangle) and \#5274 (point with errorbars) on the CMD of M4.
          All magnitudes are given in the ACS/VEGA system. The F658N filter is centered on the 
          H$\alpha$ line; the F606W magnitudes are interpolated to the phases at which frames 
          in filters F814W, F435W and F658N were taken (0.666, 0.513 and 0.073, respectively). 
          The F606W-F814W CMD is taken directly from \citet{and08}, and the remaining two CMDs 
          are extracted from subfields of the images mentioned in Sect. \ref{sect:hst}.
         }
\label{fig:cmd}
\end{figure}

The HST archive contains four images taken with the F814W filter between 05:27 and 08:04 UT 
on 2006-03-05 with $T_\mathrm{exp}=40$~s (Program ID 10775, PI A. Sarajedini). From them, 
we extracted the photometry of both 
components of the pair and checked which one was variable. The star \#5273 was constant to 
within observational errors (amounting to ~0.015~mag), while \#5274 exhibited fluctuations
with an amplitude of $\sim$0.5 mag. The final proof that the fainter component of the 
pair is responsible for the variability detected in the ground-based data came from the 
analysis of a yet another set of 30 archival $HST/ACS$ images, collected 
with the F435W filter on 2006-04-08 (Program ID 10615, PI S. Anderson; all but three of these 
obtained with $T_\mathrm{exp}=340$~s). We extracted the photometry with Dophot/Allstar 
\citep{ste87}, and transformed it to the ACS-VEGA system, following the prescription of 
\citet{sir05}. Again, the brighter star was remarkably constant ($\sigma_\mathrm{F435W} 
=0.013$ mag), showing no periodicity with $P_\mathrm{1}$ or $P_\mathrm{2}$ (in fact, it 
showed no periodicity whatsoever).

Based on $V$-magnitudes given by \citet{and08}, we estimated the instrumental 
magnitude of the star \#5274 in our ground-based reference image used for image subtraction. This  
allowed us to transform the ISIS light curve from differential counts to magnitudes (the curve 
was extracted at the position of star \#5274, and the phase difference between ISIS template 
and ACS F606W image was accounted for). Next, we phased the $V$-band magnitudes of the star 
\#5274 using the ephemeris (\ref{eq:ephem}), averaged them using phase 
bins $\Delta\phi=0.02$, and interpolated them onto phases at which the F435W frames were taken. 
The interpolation enabled us to calculate the ($V-\mathrm{F435W}$) color as a function of phase.
We found that the maxima of the light curve were bluer by $\sim$0.2 mag than the 
minima (see Fig.~\ref{fig:opt_lc}). Such a behaviour is indicative of the irradiation effect, 
which produces a light curve with one maximum per orbital period, as is observed when 
$P_\mathrm{1}$ is used phasing. Alternatively, $P_\mathrm{2}$ produces two maxima per 
period -- a feature characteristic of the ellipsoidal effect (however, in this case the 
irradiation may also play a role). In Sect.~\ref{sect:discus} we
argue that the actual orbital period $P_\mathrm{orb}$ is most probably equal to 
$P_\mathrm{1}$ ($\approx0.2628$ d $\approx6.3$ h). 

\subsection{Chandra}
\label{sect:chandr}

M4 was observed three times with the ACIS-S spectrometer on the board of 
Chandra X-ray Observatory on 2000 June 30 (Obs. ID 946), 2007 July 06
(Obs. ID 7447), and 2007 September 18 (Obs. ID 7446) for 7.17, 12.63 and 
13.31 h, respectively. During all observations the target was centered 
on the S3 chip, and the integration time was
3.24 sec per frame in the TIMED mode. All data were downloaded from the 
Chandra archive. The observations from 2000 June 30 were analyzed 
earlier by BPH, but we decided to repeat the analysis so 
that all data sets would be processed in exactly the same way. 
For the calibration of each observation the same version of CALDB 3.4.0 
was employed.

CX1 was localized using the physical coordinates given by BPH
(i.e. $\alpha = 16^\mathrm{h}23^\mathrm{m}34.12^\mathrm{s}$ and
$\delta = -26^\circ31\arcmin34.85\arcsec$), and the light curves were 
exctracted using CIAO 4.3 software. All data sets were filtered for 
energies 0.5-7 keV, as suggested in CIAO ACIS analysis threads. Background 
flaring was accounted for with the help of CIAO tools {\it dmextract} and 
{\it lc\_clean}. Since the background was practically not flaring, good-time 
intervals did not differ much from total exposure times, amounting 
to 6.6, 12.06, and 12.96 h for observations 946, 7447, and 7446, respectively.

\begin{figure}[t!]
\centering
\includegraphics[width=0.46\textwidth, angle=0, bb = 173 285 564 690,clip]{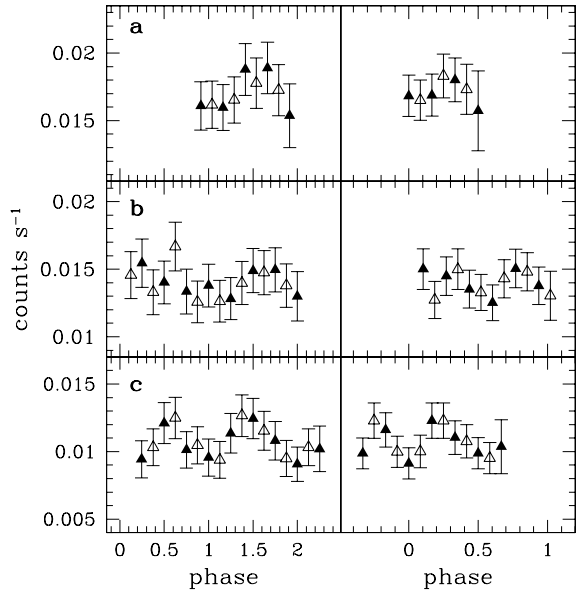}
\caption {X-ray light curves of CX 1. Left column: counts phased with 
          the epehmeris (\ref{eq:ephem}) without phase-folding, and binned 
          in four intervals per period (filled triangles). Empty triangles: 
          the same binning, but the bins are shifted by $\Delta\phi= 0.125$. 
          Right column: counts phased with $P_\mathrm{2}$ used in eq. (\ref{eq:ephem})
          instead of $P_\mathrm{1}$, and binned in six intervals per period 
          (filled triangles). Empty triangles: the same binning, but the bins are shifted 
          by $\Delta\phi= 0.083$. No phase-folding was applied.  
          Panels a, b and c show
          observations from 2000-06-30, 2007-07-06 and 2007-09-18, respectively. 
          $\phi$ = 0 corresponds to the minimum of the optical light curve.
         }
\label{fig:xray_lc}
\end{figure}

The source region was defined as a circle with a radius of 3{\arcsec} centered 
on CX 1, and the background was sampled in a surrounding annulus with inner 
and outer radius of 3\arcsec and 10\arcsec (we checked that varying the outer 
radius within reasonable limits did not change the shape of the extracted light curves). 
Data sets 946, 7447, and 7446 yielded, $410\pm20$, $610\pm24$, 
and $494\pm22$ counts, respectively. The counts were binned in intervals of 0.25
and 0.167 of the period for $P_\mathrm{1}$ and $P_\mathrm{2}$, respectively, and
the resulting X-ray light curves are shown in Fig.~\ref{fig:xray_lc}. While  
rather noisy, they exhibit a modulation similar to that of the optical light 
curves, with a minimum near $\phi=0$. 

The same regions were used for the extraction of spectra with the 
{\it specextract} tool. To within standard errors, in each data set the same 
number of counts was detected as before. The spectra were binned,
and each bin contained at least 15 counts. XSPEC v.12.6.0 was used for fitting. 
We tried to fit three models: a power law, 
a black-body, and thermal bremsstrahlung, each convolved with a warm 
absorption model. For all three data sets the only reasonable fit was 
the power law. The black-body 
produced either $r\chi^2/\mathrm{dof} > 2 $, or warm absorption dropping to 
zero, whereas thermal bremstralung required an unacceptably high temperature (199 keV).

The power law fit to the 946 data set yielded an absorbing column density 
$N_H = 2.29 \pm 1.04\times 10^{21}$ cm$^{-2}$, and photon index
$\Gamma = 0.96 \pm 0.17$ ($\chi^2/\mathrm{dof}=25.5/22=1.16$),
which  agrees well with the best-fit 
results of BPH, who obtained $N_H = (1.95-2.86)\times 10^{21}$ cm$^{-2}$ 
and $\Gamma = 0.99 \pm 0.17$ ($\chi^2/\mathrm{dof}=33.3/39=0.85$). The absorbing column 
density derived from fitting also agrees with $N_H = 2.18\times 10^{21}$, converted from the 
visual extinction $A_V=1.22$ (Kaluzny et al. 2012, in preparation) according to 
the prescription of \citet{pred95}. Assuming a distance to M4 of 2.2 kpc \citep{har96}, 
we derived an unabsorbed X-ray luminosity 
$L_{0.5-2.5\mathrm{keV}} = 3.4\times10^{31}$ erg/s, and a hardness ratio 
$L_{0.5-1.5\mathrm{keV}}/L_{1.5-6.0 \mathrm{keV}} = 0.21$. Similar values were obtained
for the remaining two observations.

\begin{figure}[t!]
\begin{center}
\includegraphics[width=0.46\textwidth, angle=0, bb = 24 5 772 524,clip]{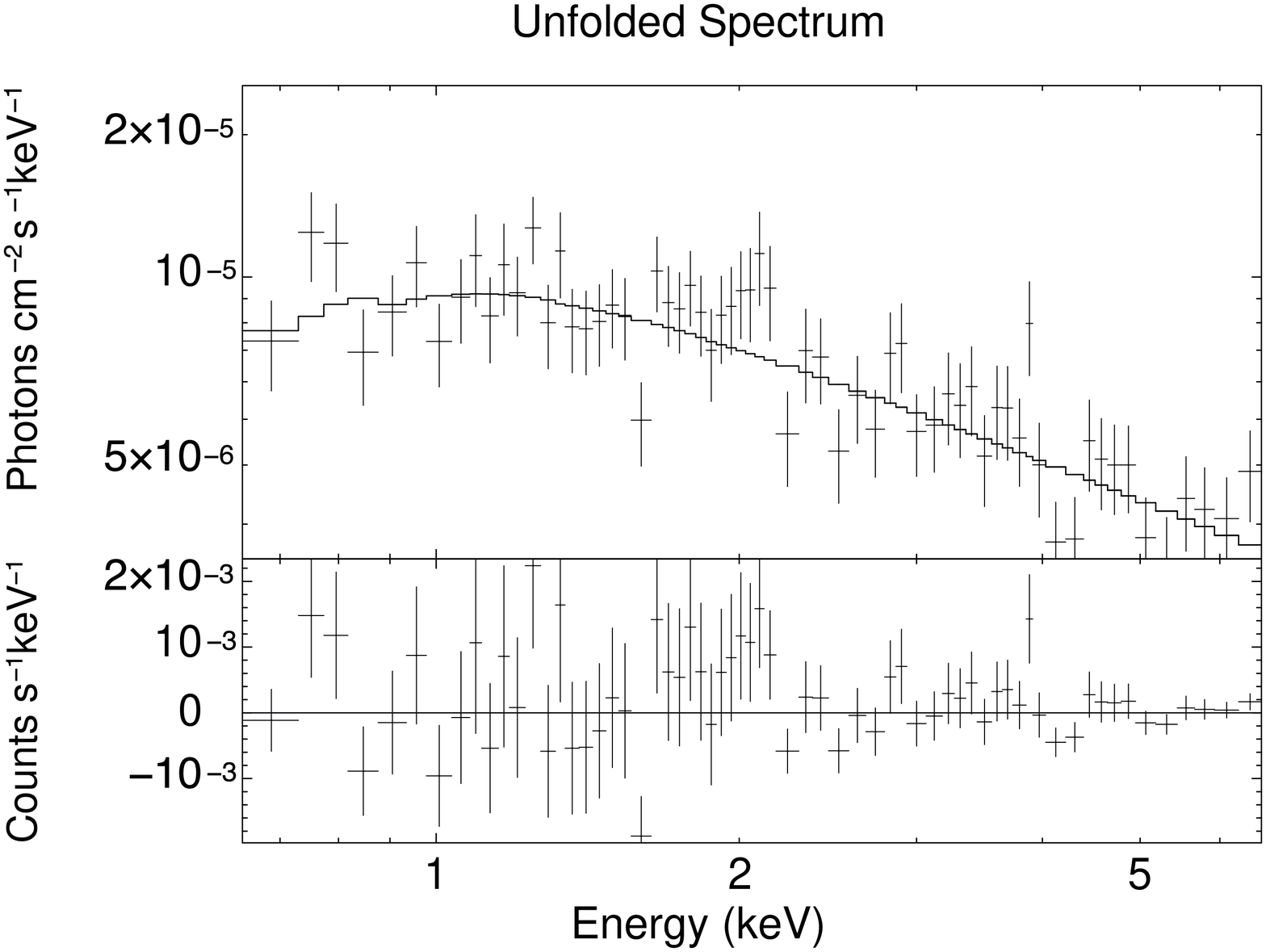}
\caption {X-ray spectrum of CX 1, derived from the combined exposures described in the text.
          The lower panel shows the residual to the power law fit. 
         }
\label{fig:xray_sp}
\end{center}
\end{figure}

To improve the photon statistics, we combined all spectral data using the CIAO tool 
{\it combine\_spectra}. The resulting cumulative spectrum of 1520 counts
was binned in such a way that each bin contained at least 20 counts. The best-fitting
model was again a power law convolved with warm absorption (Fig.~\ref{fig:xray_sp}), and its 
parameters did not differ much from those characterizing the individual spectra. We obtained
$N_H = 1.71 \pm 0.4\times 10^{21}$ cm$^{-2}$, and $\Gamma= 0.87 \pm 0.08$
($\chi^2/\mathrm{dof} =67.84/61= 1.11$), with $L_{0.5-2.5\mathrm{keV}}$ and 
$L_{0.5-1.5\mathrm{keV}}/L_{1.5-6.0 \mathrm{keV}}$ amounting to 
$2.66\times 10^{31}$ erg/s and 0.19, respectively


Our fit is significantly harder than that of BPH, resulting in a smaller
hardness ratio. Their ratio is close to unity, but it was computed using correction 
factors averaged over three fits, whereas ours is derived from the best fit to the combined 
spectrum obtained from a much larger number of counts.

\section{Discussion and conclusions} 
\label{sect:discus}

According to \citet{zlocz12}, the star \#5273 is a proper-motion member of M4.
Upon averaging the subtracted ISIS-images on a seasonal basis, we found no evidence 
of bipolar residuals at the position of its weak companion. Such residuals are 
observed for objects with noticeable proper motions with respect to the 
surrounding stellar field \citep{eye01}. Given the time base and the pixel scale of 
our ground-based images, we rule out a relative proper motion of the star \#5274 in 
excess of 10 mas/y. Since on the proper-motion diagram most of the field stars 
are separated from the cluster members by 15 mas/y \citep{zlocz12}, this is consistent with 
the star \#5274 being a member~of~M4. 

BPH tentatively classified CX 1 as a CV. However, between 1995 and 
2008 average optical luminosity and amplitude of the star \#5274 remained constant to better 
than 0.05 mag. This would be very unusual for a CV whose substantial X-ray emission implies
active accretion via L1 overflow. If CX~1 is indeed such a system, despite its stability, then its 
visible component should be a main-sequence star filling its Roche lobe, whose
average density can be calculated from the formula 
\begin{equation}
  \bar{\rho}~\mathrm{[g}~\mathrm{cm}^{-3}\mathrm{]} = 107 P_\mathrm{orb}^{-2}\mathrm~{[h]},
  \label{eq:mrho}
\end{equation}
accurate to 3\% for mass ratios $0.01<q<1.0$ \citep{eg83}. For $P_\mathrm{1}$ $(\approx$0.2628 d)
and $P_\mathrm{2}$ $(\approx$0.5256 d) we get a $\bar{\rho}$ of 2.8 and 0.7~${\rm g\ cm^{-3}}$, 
respectively.
Since an 11~Gyr old main-sequence star with $M_{\rm V}=7.15$ and ${\rm [Fe/H]}=-0.7$ should 
have $\bar{\rho}=4.04$~g~cm$^{-3}$ \citep{dotter08}, the orbital period of CX 1 
cannot be equal to $P_\mathrm{2}$. $P_\mathrm{1}$ might be considered marginally acceptable, 
were it not for the extraordinary stability of the star \#5274. 

Another argument against CX 1 being a cataclysmic variable follows from 
the ground-based UBV photometry of M4 published by \citet{moc02}. Between the outbursts, 
cataclysmic variables typically have $(U-V)\approx-1$ \citep{war95}. In the data of \citet{moc02} 
the stars \#5273 and \#5274 are blended, and the $(U-V)$ index of the unresolved blend amounts to 
1.016, locating it right in the middle of the upper main sequence in the $V/U-V$ diagram.
With $(U-V)=-1$ (-0.6), the star \#5274 would have an $U$ of 19.55 (19.95) mag, and the color of
the blend would be shifted $\delta(U-V) = -0.28$ (-0.20). Such shifts would be easily seen on 
the $V/(U-V)$ diagram of \citep{moc02}.

We conclude that mean density, stability and color arguments taken together make the interpretation 
of CX 1 as a cataclysmic variable very unlikely. 
The only viable alternative is a system whose primary is a neutron star instead of a white 
dwarf. If $P_\mathrm{orb}=P_\mathrm{1}$ then the optical variability of CX~1 must originate 
predominantly from irradiation (cf. Sect \ref{sect:hst}). However, even in this case the 
star \#5274 is likely to nearly fill its Roche lobe (first - it is heated, and therefore swollen; 
second - there must be a Roche-lobe overflow for the accretion to go on, and the X-rays to be produced).
Alternatively, if $P_\mathrm{orb}=P_\mathrm{2}$ then the ellipsoidal effect becomes the main cause 
of optical variations. In any case, the semi-detached configuration should be a reasonable 
approximation of the CX 1 system, and the estimates of the mean density of the star \#5274 based on 
equation (\ref{eq:mrho}) remain valid. Having the same luminosity and temperature as similarly 
located main-sequence stars of M4, the star \#5274 must also have the same radius 
$R\approx$0.6${R_\odot}$ \citep{dotter08}. For that size 
one gets a mass of $\sim$0.4${M_\odot}$ and $\sim$0.1${M_\odot}$, respectively for 
$P_\mathrm{1}$ and $P_\mathrm{2}$. Since the corresponding main-sequence mass is 0.58 ${M_\odot}$,
the Roche-lobe filling companion of CX 1 would have to be ablated either slightly 
(if $P_\mathrm{orb}=P_\mathrm{1}$) or very strongly (if $P_\mathrm{orb}=P_\mathrm{2}$).
Simple experiments with the $PHOEBE$ tool \citep{prsa05} show that the $V$-light curve for 
$P_\mathrm{1}$ can be easily recovered, e.g. in a system with neutron star of 1.5$M_\sun$, 
semimajor axis of 2.3$R_\sun$, inclination of $\sim$20$^\circ$ and illuminated-area temperature 
20\% higher than the rest of the photosphere (obviously, this does not prove that $P_\mathrm{1}$ is
the actual orbital period, nor the semi-detached 
configuration is the only possible).

For $P_\mathrm{orb}=P_\mathrm{1}$ there is only one X-ray minimum per period, which can be naturally 
explained as an eclipse, reminiscent of those observed in systems containing millisecond pulsars 
J0024-7203W in 47 Tuc or J1740-5340A in NGC 6397 \citep{hei11}. Two X-ray mimima per period, appearing 
for $P_\mathrm{orb}=P_\mathrm{2}$ (see Fig. \ref{fig:xray_lc}), would be very hard to explain. 
We conclude that, while $P_\mathrm{2}$ cannot be 
ruled out, it seems much less likely than $P_\mathrm{1}$. 

Thus, the available data favor the possibility that CX 1 is composed of a neutron star accompanied 
by a $\sim$0.6${M_\odot}$ main-sequence star or a partially ablated, slightly oversized 
$\sim$0.4${M_\odot}$ star. If this interpretation is correct, then it must be a millisecond pulsar or 
a qLMXB. The X-ray luminosity and hardness ratio derived in Sect. \ref{sect:chandr} support 
the first 
possibility; in fact, they locate CX 1 among the brightest millisecond 
pulsars shown in Fig. 6 of \citet{bog10} (the nondetection of the radio emission may be due to an 
unfavourable orientation). As for the radiation mechanism, 
the strongly nonthermal spectrum suggests that the dominant source of X-rays may be a relativistic 
shock, produced by the pulsar wind interacting with the matter from the optical companion 
\citep{bog05}. Future observations with XMM-Newton (more detailed X-ray light curves), HST 
(multicolor photometry) and VLT or Gemini South (IR spectroscopy with adaptive optics) should 
resolve the period ambiguity and more precisely establish the nature of this interesting object.

\acknowledgments

This research has made use of the software provided by the Chandra X-Ray Center
in the application package CIAO.
JK and MR were supported by the grant N N203 379936 from the Polish Ministry 
of Science and Higher Education. We thank the anonymous referee for helpful comments, and in 
particular for noting an error in the phasing of the F435 light curve. 

Facilities: \facility{HST}, \facility{CXO}


\begin{thebibliography}{}
\bibitem[\protect\citeauthoryear{Alard \& Lupton}{1998}]{ala98}
 Alard, C., Lupton, R.~H 1998, \apj, 503, 325
\bibitem[\protect\citeauthoryear{Anderson et al.}{2008}]{and08}
 Anderson, J. et al. 2008, \aj, 135, 2055
\bibitem[\protect\citeauthoryear{Bassa et al.}{2004}]{bas04}
 Bassa, C. et al. 2004, \apj, 609, 755
\bibitem[\protect\citeauthoryear{Bogdanov et al.}{2010}]{bog10}
 Bogdanov, S. et al. 2010, \apj, 709, 241
\bibitem[\protect\citeauthoryear{Bogdanov et al.}{2005}]{bog05}
 Bogdanov, S., Grindlay, J. E., \& van den Berg, M. 2005, \apj, 630, 1029
\bibitem[\protect\citeauthoryear{Dotter et al.}{2008}]{dotter08}
 Dotter, A. et al. 2008, \apjs, 178, 89
\bibitem[\protect\citeauthoryear{Eggleton}{1983}]{eg83}
 Eggleton, P. P. 1983, \apj, 268, 368 
\bibitem[\protect\citeauthoryear{Eyer \& Wozniak}{2001}]{eye01}
 Eyer, L., \& Wo\'zniak, P. R 2001, \mnras, 327, 601
\bibitem[\protect\citeauthoryear{Harris}{1996}]{har96}
 Harris, W. E. 1996, \aj, 112, 1487\\ 
 (www.physics.mcmaster.ca/Globular.html)
\bibitem[\protect\citeauthoryear{Heinke}{2011}]{hei11}
 Heinke, C. O. 2011, arXiv:1101.5356
\bibitem[\protect\citeauthoryear{Kaluzny et al.}{2005}]{jka05}
 Kaluzny, J. et al. 2005, AIP Conf. Proc., Vol. 752, 
 "Stellar Astrophysics with the World's Largest Telescopes", 
 ed. J. Mikolajewska and A. Olech, p. 70
\bibitem[\protect\citeauthoryear{Kaluzny et al.}{2010}]{jka10}
 Kaluzny, J., Thompson, I. B., Krzeminski, W., Zloczewski, K. 2010,
 Acta Astron., 60, 245 
\bibitem[\protect\citeauthoryear{Mochejska et al.}{2002}]{moc02}
 Mochejska, B., Kaluzny, J., Thompson, I., \& Pych, W. 2002, 
 \aj, 124, 1486
\bibitem[\protect\citeauthoryear{Pooley}{2010}]{poo10}
 Pooley, D. 2010, Proc. Natl. Acad. Sci., 107, 7164
\bibitem[\protect\citeauthoryear{Predehl \& Schmitt}{1995}]{pred95}
 Predehl, P., \& Schmitt, J. H. M. M. 1995, \aap, 293, 889
\bibitem[\protect\citeauthoryear{Pr\v sa \& Zwitter}{2005}]{prsa05}
 Pr\v sa, A., \& Zwitter, T. 2005, \aj, 628, 426
\bibitem[\protect\citeauthoryear{Schwarzenberg-Czerny}{1996}]{asc96}
 Schwarzenberg-Czerny A. 1996 \apj, 460, L107
\bibitem[\protect\citeauthoryear{Sirianni et al.}{2005}]{sir05}
 Sirianni, M. et al.  2005, \pasp, 117, 1049
\bibitem[\protect\citeauthoryear{Stetson}{1987}]{ste87}
 Stetson, P. B. 1987, \pasp, 99, 191
\bibitem[\protect\citeauthoryear{Warner}{1995}]{war95}
 Warner, B. 1995, Cataclysmic Variable Stars 
 (Cambridge: Cambridge University Press)
\bibitem[\protect\citeauthoryear{Zloczewski}{2012}]{zlocz12}
 Zloczewski, K. 2012, PhD thesis, NCAC
\end{thebibliography}
\end{document}